\documentclass[dvips]{article}
\usepackage{icrctc07}

\title{Theory of nonlinear particle acceleration at shocks and self-generation of
the magnetic field}
\shorttitle{Nonlinear theory of shock acceleration}

\authors{Pasquale Blasi$^{1}$, Elena Amato$^{1}$}
\shortauthors{Blasi and Amato}
\afiliations{$^1$INAF/Osservatorio Astrofisico di Arcetri\\
Largo E. Fermi 5, 50126 Firenze (Italy)}
\email{blasi@arcetri.astro.it}

\abstract{We present some recent developments in the theory of
  particle acceleration at shock fronts in the presence of dynamical
  reaction of the accelerated particles and self-generation of
  magnetic field due to streaming instability. The spectra of
  accelerated particles, the velocity, magnetic field and temperature
  profiles can be calculated in this approach anywhere in the
  precursor and in the downstream region. The implications for the
  origin of cosmic rays and for the phenomenology of supernova
  remnants will be discussed.}

\begin{document}
\maketitle
\section{Introduction}

The paradigm of supernova remnants as sources of the galactic component
of cosmic rays heavily relies on the fact that the mechanism of particle
acceleration at collisionless shocks should take place in the nonlinear
regime. The nonlinearity manifests itself in at least two respects: 1)
the efficiency of acceleration necessary to explain the origin of
cosmic rays needs to be large enough that the dynamical reaction of
the accelerated particles is not negligible; 2) the accelerated
particles are responsible for the self-generation of the magnetic
disturbances which in turn scatter the particles thereby allowing
their acceleration to the maximum energies observed in cosmic ray
data, most notably the KASCADE data in the knee region \cite{kascade}. 

A special emphasis should be given to the fact that protons may be
energized to energies of $10^5-10^6$ GeV only if the magnetic field in 
the shock vicinity is amplified by a factor few hundreds and
reorganized in the form of a flat power spectrum which may lead to
Bohm diffusion. Such amplification may take place through streaming
instability induced by the super-Alfvenic drift of accelerated
particles in the frame of the plasma upstream of the shock
\cite{bell78a,lagage83a,bell04}. Magnetic field amplification may also
take place due to firehose instability \cite{firehose}. 

A comprehensive theory of particle acceleration in the nonlinear
regime, with both dynamical reaction of the accelerated particles and
magnetic amplification taken into account has been recently formulated
in \cite{amato1,amato2}. In \cite{caprioli} the calculation of the
maximum momentum for the nonlinear regime was presented. 

Here we illustrate the basic formalism and the phenomenological
implications of the theoretical approach of \cite{amato2}.

\section{The formalism}

The two basic equations needed in this section are the equation of
conservation of momentum and the transport equation for the
accelerated particles. In the upstream plasma, conservation of
momentum reads:
\begin{equation}
\xi_c (x) = 1 + \frac{1}{\gamma_g M_0^2} - U(x) - \frac{1}{\gamma_g M_0^2}
U(x)^{-\gamma_g},
\label{eq:normalized1}
\end{equation}
where $\xi_c (x) = P_{CR}(x)/\rho_0 u_0^2$ and $U(x)=u(x)/u_0$ and we
used conservation of mass $\rho_0 u_0 = \rho(x) u(x)$ (here $\rho_0$
and $u_0$ refer to the density and plasma velocity at upstream infinity, 
while $\rho(x)$ and $u(x)$ are the density and velocity at the location 
$x$ upstream. $M_0$ is the sonic Mach number at upstream infinity). 
The pressure in the form of accelerated particles is defined as
\begin{equation}
P_{CR}(x) = \frac{1}{3} \int_{p_{inj}}^{p_{max}} dp\ 4 \pi p^3 v(p) f(x,p),
\end{equation}
and $f(x,p)$ is the distribution function of accelerated particles. 
Here $p_{inj}$ and $p_{max}$ are the injection and maximum momentum. The 
function $f$ vanishes at upstream infinity. The distribution function 
satisfies the following transport equation in the reference frame of
the shock: 
\begin{eqnarray*}
\frac{\partial}{\partial x}
\left[ D(x,p)  \frac{\partial}{\partial x} f(x,p) \right] - 
u  \frac{\partial f (x,p)}{\partial x} +\\ 
\frac{1}{3} \left(\frac{d u}{d x}\right)
~p~\frac{\partial f(x,p)}{\partial p} + Q(x,p) = 0.
\label{eq:trans}
\end{eqnarray*}
\cite{amato1} and \cite{malkov} showed that an excellent approximation
to the solution $f(x,p)$ has the form 
\begin{equation}
f(x,p) = f_0(p) \exp\left[-\frac{q(p)}{3}\int_x^0 dx' \frac{u(x')}{D(x',p)}
\right],
\label{eq:solution}
\end{equation}
where $f_0(p)=f(x=0,p)$ is the cosmic rays' distribution function at the shock 
and $q(p)=-\frac{d\ln f_0(p)}{d \ln p}$ is its local slope in momentum
space.

The function $f_0(p)$ can be written in a very general way as found by
\cite{blasi1}:
\begin{eqnarray*}
f_0 (p) = \left(\frac{3 R_{tot}}{R_{tot} U_p(p) - 1}\right) 
\frac{\eta n_0}{4\pi p_{inj}^3}\times \\
\exp \left\{-\int_{p_{inj}}^p 
\frac{dp'}{p'} \frac{3R_{tot}U_p(p')}{R_{tot} U_p(p') - 1}\right\}.
\label{eq:inje}
\end{eqnarray*}
Here we introduced the function $U_p(p)=u_p/u_0$, with
\begin{equation}
u_p = u_1 - \frac{1}{f_0(p)} 
\int_{-\infty}^0 dx (du/dx)f(x,p)\ ,
\label{eq:up}
\end{equation}
where $u_1$ is the fluid velocity immediately upstream (at $x=0^-$).
We used $Q(x,p) = \frac{\eta n_{gas,1} u_1}{4\pi p_{inj}^2} 
\delta(p-p_{inj})\delta(x)$, with $n_{gas,1}=n_0 R_{tot}/R_{sub}$ the 
gas density immediately upstream ($x=0^-$) and $\eta$ the fraction of 
the particles crossing the shock which are going to take part in the 
acceleration process. In the expressions above we also introduced the
compression factor at the subshock $R_{sub}=u_1/u_2$ and the total
compression factor $R_{tot}=u_0/u_2$. If the upstream plasma only
evolves adiabatically, the two compression factors are related through
the following expression  
(\cite{blasi1}):
\begin{equation}
R_{tot} = M_0^{\frac{2}{\gamma_g+1}} \left[ 
\frac{(\gamma_g+1)R_{sub}^{\gamma_g} - (\gamma_g-1)R_{sub}^{\gamma_g+1}}{2}
\right]^{\frac{1}{\gamma_g+1}},
\label{eq:Rsub_Rtot}
\end{equation}
where $M_0$ is the Mach number of the fluid at upstream infinity and 
$\gamma_g$ is the ratio of specific heats for the fluid. The parameter
$\eta$ in Eq.~\ref{eq:inje} contains the very important information 
about the injection of particles from the thermal pool. The injection
is modelled as proposed in \cite{vannoni}:
\begin{equation}
\eta = \frac{4}{3\pi^{1/2}} (R_{sub}-1) \xi^3 e^{-\xi^2}.
\end{equation}
Here $\xi$ is a parameter that identifies the injection 
momentum as a multiple of the momentum of the thermal particles in
the downstream section ($p_{inj}=\xi p_{th,2}$). The latter is 
calculated self-consistently from the Rankine-Hugoniot relations at the
subshock. For the numerical calculations that follow we always use 
$\xi=3.5$, that corresponds to a fraction of order $10^{-4}$ of the
particles crossing the shock to be injected in the accelerator. 

The scattering properties of the background plasma are described by
the scalar function $D(p)$, the diffusion coefficient. 
Once $\mathcal{F}(x,k)$ is known, the diffusion coefficient is known in
turn (\cite{bell78a}):
\begin{equation}
D(x,p)=\frac{4}{\pi}\ \frac{r_L\ v}{3\ \mathcal{F}}\ .
\label{eq:diff}
\end{equation}
From the latter equation, where $r_L$ stands for the Larmor radius of 
particles of momentum $p$, it is clear that the diffusion coefficient tends 
to Bohm's expression for $\mathcal{F} \rightarrow 1$. The expected
saturation level for the overall energy density of the perturbed
magnetic field can be easily evaluated from the fact that
\begin{equation}
\frac{B_0^2}{8 \pi} \int \frac{dk}{k}\sigma\ \mathcal{F}(k,x)=
v_A \frac{dP_{CR}}{dx}.
\label{eq:perten}
\end{equation}
Integration of this equation is straightforward when non-linear 
effects on the fluid are neglected so that $u$ and $v_A$ are both 
spatially constant. One obtains $\delta B^2/8 \pi=(v_A/u) P_{CR}$, or, in
terms of amplification of the ambient magnetic field:
\begin{equation}
\left(\frac{\delta B}{B_0}\right)^2=2\ M_A\ \frac{P_{CR}}{\rho_0 u_0^2}\ ,
\label{eq:ampl}
\end{equation}
with $M_A=u_0/v_A$ the Alfv\'enic Mach number. 

It is worth stressing that for $P_{CR}/\rho_0 u_0^2\sim 1$ and $M_A\gg 
1$, the predicted amplification of the magnetic field exceeds unity. 
CLearly this implies that the quasi-linear theory used here loses
validity and that a more accurate description, though very difficult
to achieve should be sought.

The set of equations of conservation of momentum and transport
equation with a diffusion coefficient determined as described above
can be solved by using the iterative procedure described in detail in
\cite{amato1,amato2}. 

\section{Results}

\begin{figure}
\begin{center}
\noindent
\includegraphics [width=0.45\textwidth]{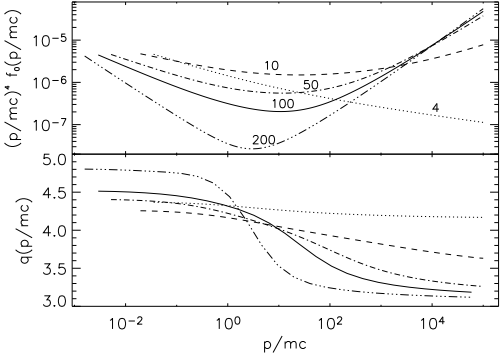}
\end{center}
\caption{Spectrum and slope at the shock location as functions of 
  energy for $p_{max}=10^5\rm m c$ and magnetic field at upstream 
  infinity $B_0=1\mu G$. The curves refer to Mach numbers at upstream 
  infinity ranging from $M_0=4$ to $M_0=200$: dotted for $M_0=4$, 
  dashed for $M_0=10$, dot-dashed for $M_0=50$, solid for $M_0=100$ 
  and dot-dot-dashed for $M_0=200$.}\label{fig:varymn}
\end{figure}

The spectra of the accelerated particles for Mach numbers at upstream
infinity ranging from $M_0=4$ to $M_0=200$ are shown in
Fig.~\ref{fig:varymn} for a background magnetic field at upstream
infinity $B_0=1\mu G$ (the result is however independent of the
strength of the background magnetic field). In the bottom part of the
same figure we plot the slope of the spectrum as a function of
momentum.  

For low Mach numbers and at given $p_{max}$ the
modification of the shock due to the reaction of the accelerated
particles is small (see for instance the case $M_0=4$). For the
strongly modified case (e.g. $M_0=200$) the asymptotic spectrum of the
accelerated particles is very flat, tending to $p^{-\alpha}$ with
$\alpha=3.1-3.2$ for $p \rightarrow p_{max}$. 

We need to comment on the issue of the shape of the spectrum of cosmic
rays accelerated in supernova remnants: the spectra illustrated in
Fig. \ref{fig:varymn} are the spectra in proximity of the shock
surface and therefore the ones which are relevant for the calculation
of the spectra of secondary radiation produced by the accelerated
particles. In general the spectra that are observed by an observer far
upstream are more complex to determine: at each time during the
supernova evolution particles at the maximum momentum can escape to 
upstream infinity (this is a peculiar aspect of nonlinear theory)
carrying away a sizeable fraction of the total energy (due to the flat
spectra). At each time the instantaneous spectrum that escapes to
upstream infinity is a narrow function centered around $p_{max}(t)$,
where $t$ is the age of the remnant. If the maximum momentum decreases
with time (in the Sedov phase this is the case) then a spectrum is
built due to the overlap of many delta-function-like spectra leaving
the acceleration region from upstream. In the simple model considered
in \cite{ptuskin} this overlap leads to a power law spectrum $p^{-4}$,
despite the fact that the spectrum at the shock may be concave as
illustrated in Fig. \ref{fig:varymn}. In addition to these particles
that leave the system from upstream, the distant observer will also
measure the cosmic ray spectrum which is kept in the supernova shell
and is eventually liberated at later times, possibly suffering adiabatic
energy losses. The spectrum observed at the Earth is likely to be a
complex superposition of the particles escaping from upstream, those
leaving the system at the end of the supernova evolution, and summed
over all supernova events occurred during a confinement time of cosmic
rays in the Galaxy.

\begin{figure}
\begin{center}
\noindent
\includegraphics [width=0.45\textwidth]{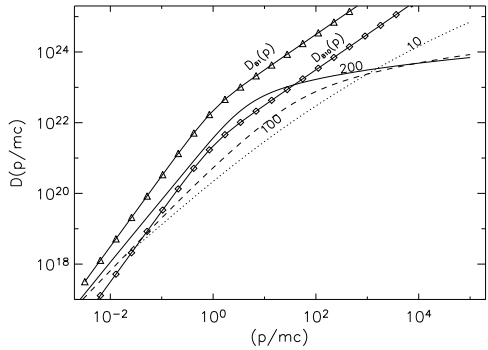}
\end{center}
\caption{The self-generated diffusion coefficient at the shock location
  $x=0^-$ as a function of the particle momentum for Mach numbers
  $M_0=10$ (dotted line), $M_0=100$ (dashed line) and $M_0=200$ (solid
  line). Also plotted is the Bohm diffusion coefficient corresponding 
  to $B_0=1 \mu G$ (solid line with triangles) and $B_0=10\mu G$
  (solid line with diamonds). The $y$-axis is in units of ${\rm cm}^2
  {\rm s}^{-1}$.}\label{fig:diff}
\end{figure}

The diffusion coefficient associated with the self-generated waves is
given by Eq.~\ref{eq:diff}. We plot this diffusion coefficient at the
shock location in Fig.~\ref{fig:diff} for Mach numbers $M_0=10$ (dashed 
lines) and $M_0=100$ (solid lines). For comparison, we also plot the
corresponding Bohm diffusion coefficient $D_B(p)\propto v(p)p$ in the
unperturbed magnetic field $B_0$, for $B_0=1\mu G$ and $B_0=10\mu G$. 

As stressed above, in the regime we considered, the fluctuations in
the magnetic field become strongly non linear. The dynamical role of
the amplified field remains however negligible as the highest values 
of $\delta B^2/8\pi\rho_0 u_0^2$, reached close to the shock front,
are of the order of $10^{-2}-10^{-3}$. 

\section{Conclusions}

We developed a mathematical formalism that allows us to calculate the
spectrum of particles accelerated at a collisionless non-relativistic
shock taking into acount both the nonlinear dynamical reaction of the
accelerated particles and the magnetic field amplification by
streaming instability. The approach has also recently been generalized
to allow us to determine self-consistently the maximum momentum
reached by the particles in this fully nonlinear acceleration regime
\cite{caprioli}.

\end{document}